\documentclass{aastex631}
\usepackage{amsmath}

\newcommand{\G}{\Gamma}

\newcommand{\p}{^\prime}
\newcommand{\e}{\epsilon}

\newcommand{\g}{\gamma}

\newcommand{\psim}{\lower.5ex\hbox{$\; \buildrel \propto \over\sim \;$}}
\newcommand{\lbar}{\lower.0ex\hbox{$\; \buildrel{\lower0.0ex \hbox{-}} \over\lambda  \;$}}

\newcommand{\m}{\mathrm{m}}
\newcommand{\cm}{\mathrm{cm}}
\newcommand{\km}{\mathrm{km}}
\newcommand{\erg}{\mathrm{erg}}

\newcommand{\s}{\mathrm{s}}
\newcommand{\dday}{\mathrm{day}}

\newcommand{\Hz}{\mathrm{Hz}}

\newcommand{\kpc}{\mathrm{kpc}}
\newcommand{\Mpc}{\mathrm{Mpc}}

\newcommand{\yr}{\mathrm{yr}}

\newcommand{\fermi}{{\em Fermi}}

\shorttitle{}
\shortauthors{Finke \& Razzaque}

\begin{document}

\title{Is Gamma-ray Burst 221009A Really a Once-in-10,000 Year Event?}

\author{Justin D.\ Finke}

\affil{U.S.\ Naval Research Laboratory, Code 7653, 4555 Overlook Ave.\ SW,
        Washington, DC,
        20375-5352, USA; \\
        justin.d.finke.civ@us.navy.mil}

\author{Soebur Razzaque}\altaffiliation[Also at the ]{Department of Physics, The George Washington University, Washington, DC 20052, USA; and National Institute for Theoretical and Computational Sciences (NITheCS), Private Bag X1, Matieland, South Africa, South Africa.}
\affil{Centre for Astro-Particle Physics (CAPP) and Department of Physics, \\
University of Johannesburg, P.O. Box 524, Auckland Park 2006, South Africa; \\ srazzaque@uj.ac.za \\}

\begin{abstract}

Gamma-ray bursts (GRBs) brighter than the GRB 221009A, the brightest yet observed, have previously been estimated to occur at a rate of 1 per 10,000 years, based on the extrapolation of the distribution of fluences of the Long GRB population.  We show that bursts this bright could instead have a rate as high as approximately one per 200 years if they are from a separate population of narrow-jet GRBs.  This population must have a maximum redshift of about $z\approx 0.38$ in order to avoid over-producing the observed rate of fainter GRBs.  We show that it will take $\ga 100$ years to confirm this new population based on observing another GRB from it with a $\g$-ray detector; observing an orphan optical afterglow from this population with Vera Rubin Observatory or an orphan radio afterglow with the Square Kilometer Array will also take similarly long times to observe, and it is unclear if they could be distinguished from the standard GRB population.  We show that the nearby narrow-jet population has more favorable energetics for producing ultra-high energy cosmic rays than standard GRBs. The rate of bursts in the Milky Way bright enough to cause mass extinctions of life on Earth from the narrow jet population is estimated to be approximately 1 per 500 Myr.  This GRB population could make life in the Milky Way less likely, with implications for future searches for life on exoplanets.

\end{abstract}

\section{Introduction}
\label{intro}

For brief periods of time, gamma-ray bursts (GRBs) can outshine everything else in the $\g$-ray sky.  They are generally thought to have two progenitor channels \citep{kouveliotou93}.  One is the merging of two neutron stars, which has been confirmed with the detection of gravitational waves coincident with GRB 170817A \citep{abbott17_mm,goldstein17}. These typically produce 90\% of their emission in $\la 2$\ s \citep[although there may be some exceptions; e.g.,][]{dichiara23} and consequently are referred to as Short GRBs.  A Short GRB might also be produced by the merging of a neutron star and black hole, although as yet this has not been confirmed.  Another channel for producing GRBs is core-collapse supernovae from stars stripped of their outer hydrogen atmospheres, known as Type Ib/Ic supernovae \citep[e.g.,][]{pian00,soderberg04}.  These typically produce 90\% of their emission in $\ga 2$\ s, and are referred to as Long GRBs.  Both types of progenitors produce narrow, highly relativistic jets, with bulk Lorentz factors reaching values as high as 300 or more.  Shocks within the jets accelerate particles, which then radiate $\g$-rays.  When the jets are pointed in the direction of Earth, these $\g$-rays are seen as GRBs.  After the bulk of the $\g$-rays have dissipated, the jet continues to plow into the circumburst medium, sweeping up material and slowing the jet.  During this phase the jet can radiate across the electromagnetic spectrum, from radio to $\g$-rays, creating what is known as an afterglow \citep[for a review see][]{meszaros06}.

The Long GRB 221009A was detected by almost every astrophysical $\g$-ray detector in operation, including \fermi\ Gamma-ray Burst Monitor \citep[GBM;][]{lesage23}, Konus-Wind \citep{frederiks23}, the {\em Neil Gehrels Swift Observatory} \citep{williams23}, AGILE \citep{tavani23}, INTEGRAL \citep{rodi23}, MAXI \citep{negoro22}, GRBAlpha \citep{ripa23}, Insight-HXMT, GECAM-C \citep{an23}, and SIRI-2 \citep[][L.\ Mitchell et al.\ 2024, in preparation]{mitchell22}.  It was so bright that it saturated almost every astrophysical $\g$-ray detector.  Its redshift was quickly determined to be $z=0.151$ \citep[e.g.,][]{malesani23}, quite close for a Long GRB.  It was brighter than the next brightest GRB by over an order of magnitude, both in $\g$-ray fluence and flux; and the implied isotropic-equivalent energy released ($E_{\rm iso}$) and implied isotropic-equivalent luminosity were among the brightest ever observed.  It was subsequently dubbed the ``Brightest of All Time'', or ``BOAT'' \citep{burns23}.  

Estimates of the rates for GRBs as bright as 221009A are reasons for concern.  The brightness distribution of astrophysical objects, including GRBs, are often characterized by a ``log(N)-log(S)'' plot; that is, a plot of the cumulative number of objects (N) above some fluence (S) as a function of that fluence.  The plots almost always follow a characteristic $N\propto S^{-3/2}$ at the bright end, where one finds nearby, bright objects and the Universe is approximately Euclidean.  \citet{burns23} extrapolated the log(N)-log(S) distribution for Long GRBs to the fluence for 221009A, and estimated that bursts as bright as this one should occur approximately once per 10,000 years.  It would indeed be an enormous coincidence if such a burst did appear in the $\approx 50$\ or so years since humanity has had the ability to detect GRBs; such a coincidence has an estimated probability of $\approx 50/10,000 = 0.005$ of occurring.

But perhaps there is another explanation.  For instance, it could be brighter due to gravitational lensing \citep{bloom22}.  Another possibility is that it is from a different GRB population.  Here we show that GRB 221009A could be from a different population of nearby, narrow jet Long GRBs, separate from the standard population commonly observed.  This population could make bursts such as 221009A much more common, while not over-producing the rates of fainter Long GRBs that have been observed. In Section \ref{logNlogS_section} we describe such a model.  In Section \ref{implication} we explore some of the implications of this population of GRBs, including orphan optical and radio afterglows, the origin of ultra-high energy cosmic rays, and the increased rate of extinction of life on Earth and throughout the Milky Way caused by GRBs. Finally we conclude with a discussion (Section \ref{discussion}).  We assume a flat $\Lambda$CDM Universe with cosmological parameters ($h$, $\Omega_m$, $\Omega_\Lambda$) = (0.7, 0.3, 0.7), where the Hubble constant $H_0 = 100h\ \km\ \s^{-1}\ \Mpc^{-1}$.

\section{log(N)-log(S) Model}
\label{logNlogS_section}


We follow the formalism of \citet{le07,le09,le17} to compute $N(>S)$, the number of Long GRBs observed above a given fluence $S$ using a toy model.  This distribution has also been explored by \citet{meszaros95}.  The GRB rate is given by
\begin{flalign}
\label{dotN}
\frac{d\dot N}{d\Omega dS dz} = \frac{\dot n_{\rm co}(z)}{1+z} \frac{dV_{\rm co}}{dz}  \delta(S-\hat S)  \ ,
\end{flalign}
where $z$ is the cosmological redshift of the burst, $\Omega$ is the solid angle over which one is observing, 
\begin{flalign}
\label{dotnco}
\dot n_{\rm co}(z) = K_n \psi(z)
\end{flalign}
is the rate of GRBs per co-moving volume, and $\psi(z)$ is the comoving star formation rate density.  Long GRBs have progenitors that are high-mass stars with lifetimes that are small compared to cosmological timescales, so their rate should approximately follow the star formation rate \citep[although see, e.g.,][]{yu15,petrosian15,le17}.  The normalization constant $K_n$ is the number of GRBs produced per unit solar mass of stars formed.  We use the star formation rate density parameterization from \citet{madau14}, 
\begin{flalign}
\psi(z) = a_s \frac{(1+z)^{b_s}}{1 + [(1+z)/c_s]^{d_s}}\ ,
\end{flalign}
with parameters found by \citet{finke22} from a model fit to a wide variety of luminosity density, mass density, dust extinction, and $\g$-ray extinction data. These parameters are $a_s=9.2\times10^{-3}\ M_\odot~\yr^{-1}~\Mpc^{-3}$, $b_s=2.79$, $c_s=3.10$, and $d_s=6.97$.  We expect the exact choice for the star formation rate density will have little impact on our results.  Additionally, 
\begin{flalign}
\frac{dV_{\rm co}(z)}{dz} = \frac{c d_{\rm co}(z)^2}
{H_0\sqrt{(1+z)^3\Omega_m + \Omega_\Lambda}}\ 
\end{flalign}
is the differential comoving volume element, and
\begin{flalign}
d_{\rm co}(z) = \frac{c}{H_0}\int_0^z 
\frac{dz\p}{\sqrt{(1+z\p)^3\Omega_m + \Omega_\Lambda}}
\end{flalign}
is the comoving distance.  In Equation (\ref{dotN}), $\delta$ is the usual Dirac $\delta$ function.  The fluence is given by
\begin{flalign}
\hat S = \frac{E_{\rm iso}(1+z)}{4\pi [d_L(z)]^2}, 
\end{flalign}
where
\begin{flalign}
d_L(z) = (1+z)d_{\rm co}(z)
\end{flalign}
is the luminosity distance.  The jets are assumed to have ``top-hat'' distributions of fluence, so that the energy in $\g$-rays emitted within the jet opening angle $\theta_j = \cos^{-1}\mu_j$ is
\begin{flalign}
E_{\rm \g} = E_{\rm iso}(1-\mu_j)\ .
\end{flalign}

The number of sources greater than some fluence value $S_1$--which is what is plotted in log(N)-log(S) plots---is given by
\begin{flalign}
\label{NS1}
N(>S_1) = \int_{S_1}^\infty dS N(S) = \int_{S_1}^\infty dS \int d\Omega \int dt \int_0^1 d\mu_j g(\mu_j) \int_{\mu_j}^{1} d\mu \int dE_\gamma h(E_\gamma) \int_{0}^{z_{\rm max,
*}} dz \frac{d\dot N}{d\Omega dS dz}\ ,
\end{flalign}
where $\theta = \cos^{-1}\mu$ is the angle of the jet to the line of sight; $t$ is the time of the observation; $g(\mu_j)$ is the distribution of $\mu_j$, which is normalized to unity; $h(E_\g)$ is the distribution of $E_\g$, also normalized to unity; and $z_{\rm max,*}$ is the redshift where the first GRBs are produced.  

\subsection{Rate Model with all GRB jets having the same opening angle and energy}
\label{modeldeltafcn}

Here we will make the assumption that all GRBs have the same jet opening angle, so that $g(\mu_j) = \delta(\mu_j - \hat\mu_j)$, and the same $E_\g$, so that $h(E_\g)=\delta(E_\g-\hat E_\g)$.  This assumption is not realistic, but it is a useful approximation for a simple understanding of the population.  We will explore more realistic distributions in Section \ref{distribution_section} below.  Substituting these distributions for $g(\mu_j)$ and $h(E_\g)$ and Equation (\ref{dotN}) into Equation (\ref{NS1}) and performing the integrals gives
\begin{flalign}
\label{NS2}
N(>S_1) = \Omega \Delta t K_n (1-\hat\mu_j) \int_0^{\min(z_{\rm max},z_{\rm max,*})} dz \frac{dV_{\rm co}(z)}{dz} \frac{\psi(z)}{1+z}\ ,
\end{flalign}
where $\Omega \equiv \int d\Omega$ is the solid angle the survey covers, $\Delta t \equiv \int dt$ is the length of time over which the survey takes place, and $z_{\rm max}$ can be found by solving
\begin{flalign}
\label{zmaxdef}
\frac{d_L(z_{\rm max})^2}{1+z_{\rm max}} = \frac{\hat E_\g}{4\pi S_1 (1-\hat\mu_j)}\ 
\end{flalign}
numerically.  

In the nearby Universe, $z\ll 1$, $z_{\max} \ll z_{\max,*}$, so that $d_L\approx d_{\rm co}\approx cz/H_0$, $dV_{\rm co}/dz \approx \left(c/H_0\right)^3 z^2$, Equation (\ref{zmaxdef}) gives 
\begin{flalign}
\label{zmaxapprox}
z_{\max} \approx \frac{H_0}{c\hat\theta_j} \left(\frac{\hat E_\g}{2\pi S_1 }\right)^{1/2}\ ,
\end{flalign}
and the star formation rate density reduces to $\psi(z)\approx a_s$ and becomes independent of $z$.  Then Equation (\ref{NS2}) gives
\begin{flalign}
\label{NSapprox}
N(>S_1) = \frac{\Omega \Delta t K_n a_s}{6\hat\theta_j }
\left( \frac{\hat E_\gamma}{2\pi S_1}\right)^{3/2} \ .
\end{flalign}
Thus for large $S_1$, or small $z$, $N(>S_1)\propto S_1^{-3/2}$, which is a well-known result for a Euclidean Universe.  

\begin{figure}
\vspace{2.2mm} 
\epsscale{1.1} 
\plotone{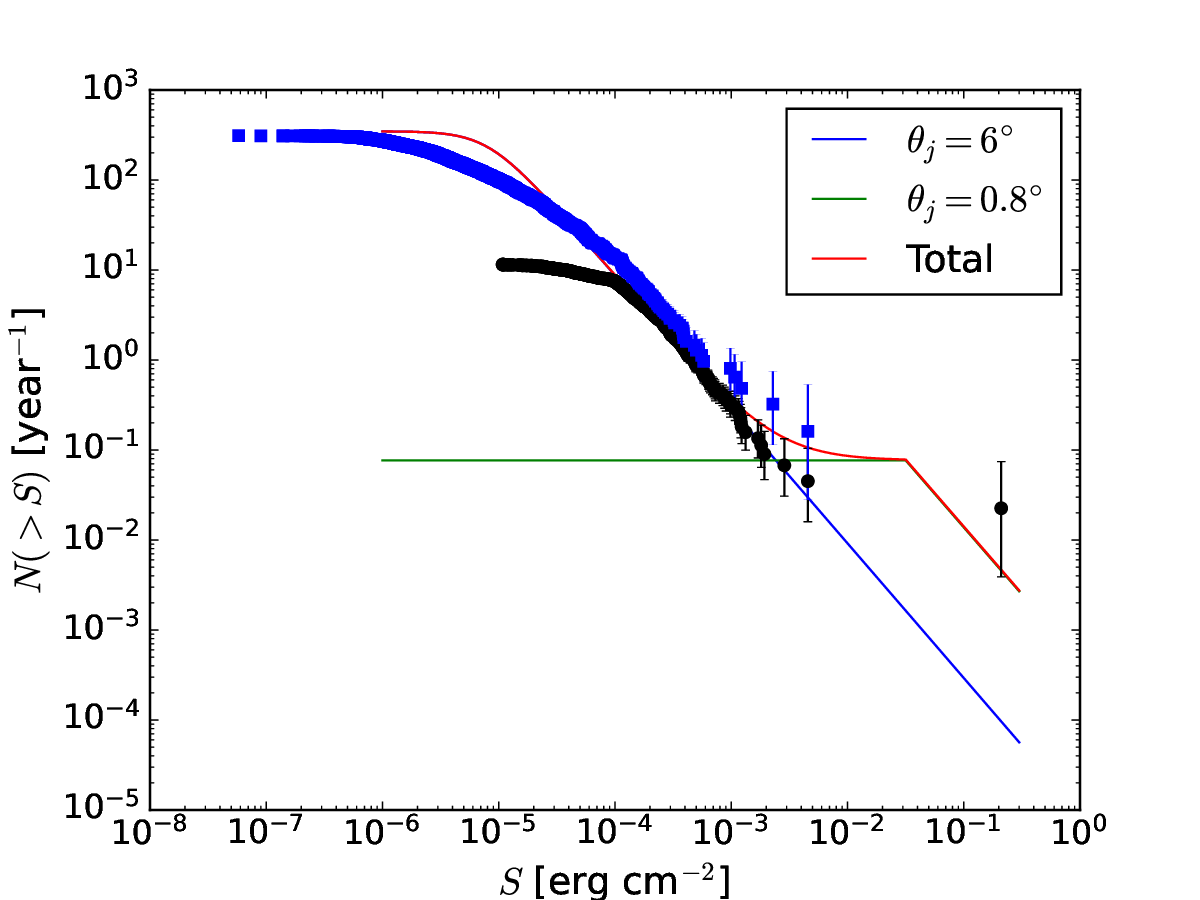}
\caption{A plot of $N(>S_1)$ versus $S_1$ (i.e., the ``log(N)-log(S)" distribution).  The blue squares and black circles are the observed GBM and total GRBs, taken from \citet{burns23}.  The curves are the $\delta$ function models (from Section \ref{modeldeltafcn}) with $E_\g=10^{51}\ \erg$ and two GRB populations with different $\theta_j$, as indicated in the legend.  The $\theta_j=6\arcdeg$ population uses $z_{\max,*}=10$ while the $\theta_j=0.8\arcdeg$ population uses $z_{\max,*}=0.36$.}
\label{logNlogS_fig1}
\vspace{2.2mm}
\end{figure}

In Figure \ref{logNlogS_fig1} we plot the log(N)-log(S) data from \citet{burns23}, including the 68\% Poisson uncertainties, computed following \citet{gehrels86}.  GRB 221009A is shown on the far right of this figure, with $S=0.21\ \erg\ \cm^{-2}$.  The distribution (excluding 221009A) follows the $\propto S_1^{-3/2}$ relation at high $S_1$, and turns over to be flat at low $S_1$.  The turnover at low fluence occurs for two reasons: (i) the GRBs become too faint for instruments to detect them, and (ii) for these low fluences, the redshifts they must come from are higher than the redshifts where these objects are produced; i.e., in our notation, here $z_{\max}$ becomes greater than  assumed $z_{\max,*}$.  

We wish to reproduce the observed population of GRBs (neglecting GRB 221009A for the time being) with our model. 
Inspecting Figure \ref{logNlogS_fig1}, in order to reproduce the GBM distribution, the fluence where the turnover from flat to $\propto S_1^{-3/2}$ occurs must be at $S_1\ga 10^{-5}\ \erg\ \cm^{-2}$.  We take $z_{\max,*}=10$, approximately the highest redshift where GRBs have been found, and $\hat E_\g=10^{51}\ \erg$, the typical value for GRBs.  This implies that, using Equation (\ref{zmaxdef}), the turnover where $z_{\max}=z_{\max,*}$ occurs must be at $\hat\theta_j\ga2.5$\arcdeg.  We take $\hat\theta_j=6\arcdeg$, which is the peak of the opening angle distribution found by \citet{goldstein16}.  Our model (Equation (\ref{NS2})) with these parameters is shown as the blue curve in Figure \ref{logNlogS_fig1}.  We adjusted the normalization so that this model matches the data from \citet{burns23}, giving $K_n=2\times10^{-5}\ M_\odot^{-1}$.  The curve nicely explains most of the GRB population for $S_1\la 10^{-2}\ \erg\ \cm^{-2}$.  For $S_1\la 2\times10^{-5}\ \erg\ \cm^{-2}$ the model diverges from the GBM data because at the faint end many GRBs are below the detectors' threshold and cannot be observed.  The ``standard'' $\theta_j=6\arcdeg$ GRB population is clearly not consistent with GRB 221009A within the 68\% Poisson uncertainties.  

We wish to explain GRB 221009A as part of a separate population of GRBs.  This second population must have $z_{\max}=0.151$ when $S_1=0.21\ \erg\ \cm^{-2}$, the measured redshift and fluence of 221009A, respectively.  From Equation (\ref{zmaxdef}), this implies that this population has
\begin{flalign}
\label{E_d_thetaj}
\frac{\hat E_\g}{\hat\theta_j^2} \approx 1.7\times10^{51}\ \erg\ \deg^{-2}\ .
\end{flalign}
The rate for GRBs brighter than 221009A must be $N(>S_1)\ga 4\times10^{-3}\ \yr^{-1}$ to be consistent with the 68\% lower limit of its Poisson uncertainty.  GRBs brighter than the next brightest burst (GRB 230307A), should occur at a rate of $N(>S_1)\la 0.1\ \yr^{-1}$ within the 68\% Poisson uncertainties.  Subtracting off the contribution from the standard GRB population gives $N_2(>S_1)\la0.07\ \yr^{-1}$, where the ``2'' subscript indicates this is only the narrow jet population.  In order for this to occur, $z_{\max,*} = z_{\max}$ at $0.1\ \yr^{-1}$.  This will occur at a fluence $S\ga 0.21\ \erg\ \cm^{-2}(0.1\ \yr^{-1}/4\times10^{-3}\ \yr^{-1})^{-2/3} \approx 0.03\ \erg\ \cm^{-2}$.    One can use this rate value to solve for where $z_{\max}=z_{\max,*}$ using Equations (\ref{zmaxdef}) and (\ref{E_d_thetaj}), giving $z_{\max,*}\la0.38$.  {\em This implies that 221009A is drawn from a separate population of GRBs which is very nearby (no farther than $z\approx0.38$).  Given our reasonable assumptions, this conclusion is unavoidable.}

GRB 221009A seems to have a narrower opening angle than most GRBs \citep[see the discussion by][]{burns23}.  We take $\hat\theta_j=0.8\arcdeg$ based on modeling its very-high energy (VHE) $\g$-ray emission observed by LHAASO \citep{cao23}. Based on Equation (\ref{E_d_thetaj}), this implies $\hat E_\g=10^{51}\ \erg$.  We assume the second population also is proportional to the star formation rate, 
\begin{flalign}
\dot n_{\rm co}(z) = K\p_n \psi(z)H(z-z_{\max,*})\ ,
\end{flalign}
with different normalization constant $K\p_n$.  Here $H(x)$ is the Heaviside function
\begin{equation}
H(x) = \begin{cases}
1 & x \ge 0 \\
0 & x < 0
\end{cases}
\ .
\end{equation}
We adjust values of $z_{\max,*}$ and $K\p_n$ in order to explain GRB 221009A and not over-produce the standard population of GRBs.  This gives $z_{\max,*}=0.36$ (consistent with the constraint described above) and $K\p_n=6\times10^{-5}\ M_\odot^{-1}$.  The result can be seen in Figure \ref{logNlogS_fig1}.  The combination of the standard, wide angle distribution, and the narrow angle distribution from which 221009A is drawn, can explain all of the relevant data within the 68\% uncertainties.


\subsection{Rate Model with distributions of energy and opening angle}
\label{distribution_section}

Now let $E_\gamma$ have some distribution $h(E_\g)$, and $\theta_j$ have some distribution $g(\mu_j)$ that are not Dirac $\delta$-functions.  Then one has the general case
\begin{flalign}
N(>S_1) = \Omega \Delta t K_n \int_{0}^{1} d\mu_j g(\mu_j)(1-\mu_j)
\int dE_\gamma h(E_\gamma) \int_0^{\min(z_{\rm max},z_{\rm max,*})} dz \frac{dV_{\rm co}(z)}{dz} \frac{\psi(z)}{1+z}\ ,
\label{logNlogS}
\end{flalign}
where now $z_{\max}$ is defined by
\begin{flalign}
\label{zmaxdef2}
\frac{d_L(z_{\rm max})^2}{1+z_{\rm max}} = \frac{E_\g}{4\pi S_1 (1-\mu_j)}\ .
\end{flalign}

\begin{figure}
\vspace{2.2mm} 
\epsscale{1.1} 
\plottwo{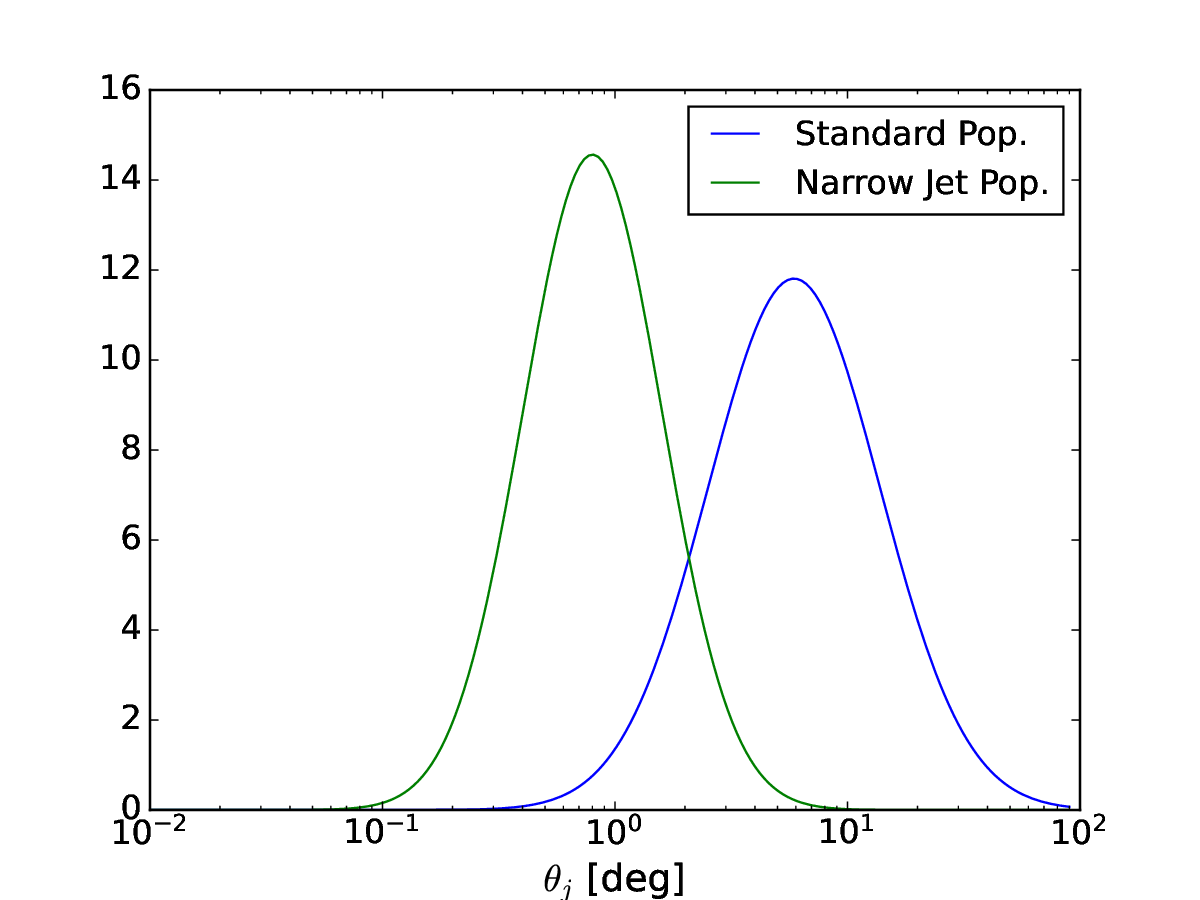}{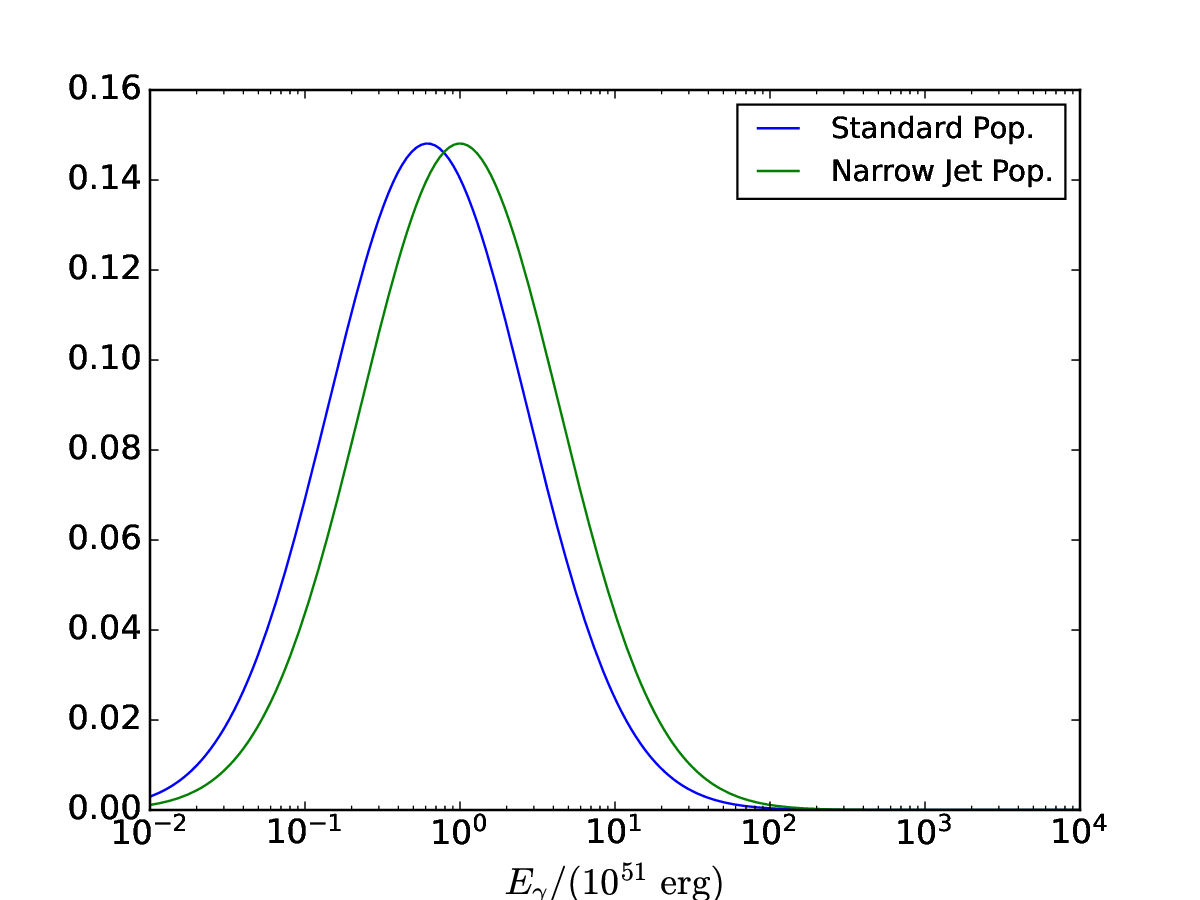}
\caption{Left:  The distributions for $\theta_j=\cos^{-1}\mu_j$ used in our calculations.  Right:  The distributions of $E_\g$ used in our calculations. }
\label{distributions_fig}
\vspace{2.2mm}
\end{figure}

\begin{figure}
\vspace{2.2mm} 
\epsscale{1.1} 
\plotone{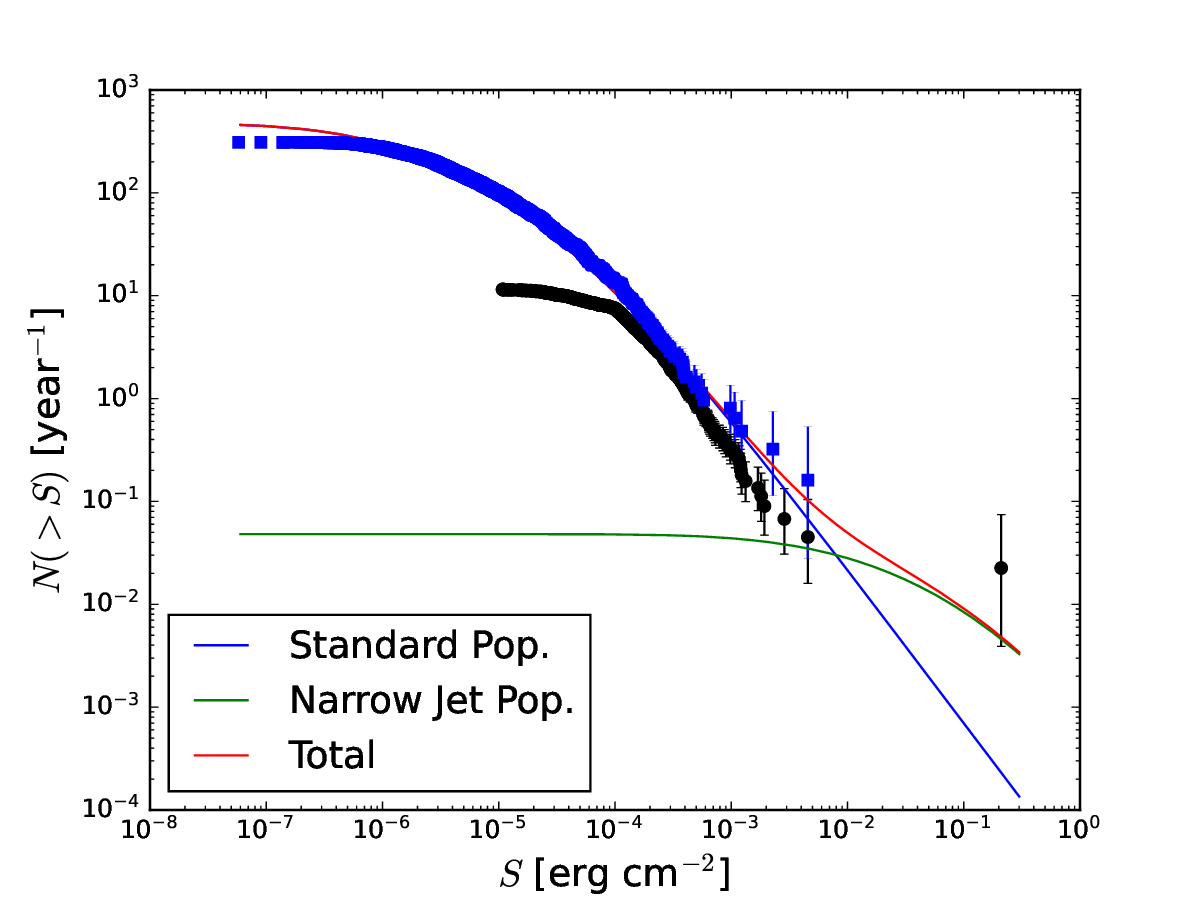}
\caption{Same as Figure \ref{logNlogS_fig1} but using
log-normal distributions for $\theta_j$ and $E_\gamma$.}
\label{logNlogS_fig2}
\vspace{2.2mm}
\end{figure}

\citet{goldstein16} estimated the distribution of $E_\g$ by using the Ghirlanda relation \citep{ghirlanda04}, a relationship between the peak energy in the GRB spectrum and $E_\g$.  A comparison of this with $E_{\rm iso}$ allowed them to also obtain the distribution of $\theta_j$.  We use the distributions from \citet{goldstein16}, which are
\begin{flalign}
\label{gmu}
g(\mu_j) = K_\theta \exp\left[\frac{-(\log_{10}\cos^{-1}\mu_j-m_\theta)^2}{2\sigma_\theta^2}\right]\ ,
\end{flalign}
and
\begin{flalign}
\label{hE}
h(E_\g) = K_\gamma \exp\left[\frac{-(\log_{10}\e_\g-m_E)^2}{2\sigma_E^2}\right]\ ,
\end{flalign}
where $\e_\g = E_\g/(10^{51}\ \erg)$.  The normalization constants $K_\theta$ and $K_\gamma$ are found by taking these distributions normalized to unity.  For our standard GRB population, we use the values from \citet{goldstein16} in these distributions, which are $m_\theta=0.77$, $\sigma_\theta=0.37$, $m_E=-0.21$, and $\sigma_E=0.64$.  For the narrow jet GRB population, we use $m_\theta=-0.097$, $\sigma_\theta=0.3$, $m_E=0$, and $\sigma_E=0.64$.  These distributions can be seen in Figure \ref{distributions_fig}.  For the standard angle jet population we use $z_{\max,*}=10$, and for the narrow jet population we use $z_{\max,*}=0.36$.  The normalization constants in Equation~(\ref{logNlogS}) were adjusted to match the distributions observed by GBM and all instruments, as found by \citet{burns23}, giving $K_n=10^{-6}\ M_\odot^{-1}$ for the standard distribution, and $K\p_n=4\times10^{-6}\ M_\odot^{-1}$ for the narrow jet distribution.  The resulting log(N)-log(S) distribution can be seen in Fig.\ \ref{logNlogS_fig2}.  

\begin{figure}
\vspace{2.2mm} 
\epsscale{1.1} 
\plotone{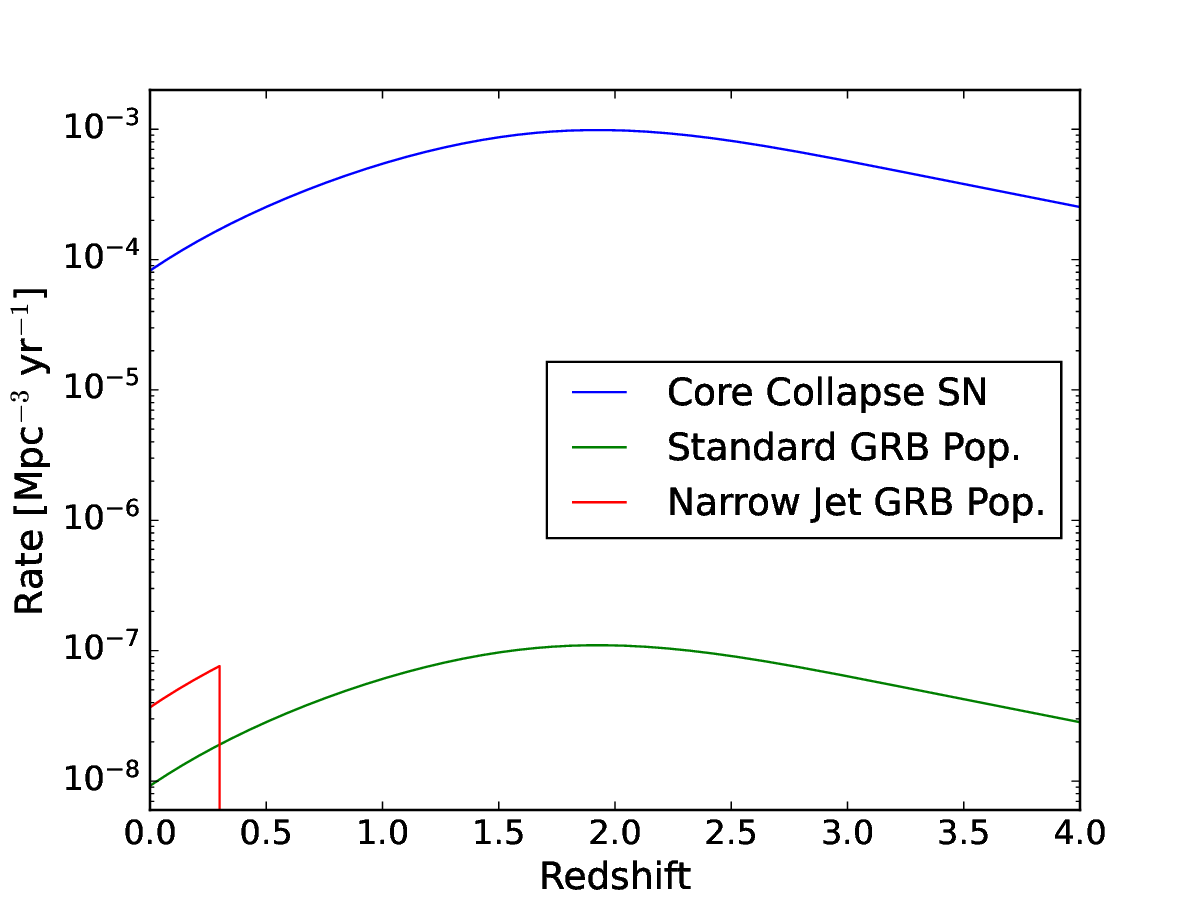}
\caption{A plot of rates per comoving volume for core-collapse and the two GRB populations in all directions (not just those pointed at the Earth).}
\label{logNlogS_rates}
\vspace{2.2mm}
\end{figure}

It is interesting to compare the rates of the GRB populations in all directions with the rate of core-collapse supernovae (CCSNe).  Since CCSNe are thought to be the parent population of long GRBs, the GRB rates should not exceed the CCSNe rate.  The core-collapse supernova rate can be estimated from \citep[e.g.,][]{hopkins06}
\begin{flalign}
\dot n_{\rm co,SNe}(z) = \psi(z) \frac{\int_8^{50} dm\ \xi(m)}{\int_{0.1}^{100} dm\ m\ \xi(m)}\ 
\end{flalign}
where $m$ is the stellar mass in solar masses and $\xi(m)$ is the stellar initial mass function (IMF).  We use the IMF as found by a fit to a wide variety of luminosity density data by \citet{baldry03}, which is
\begin{equation}
\xi(m) = N_0
\begin{cases}
(m/m_0)^{-1.5} & m \le m_0 \\
(m/m_0)^{-2.2} & m > m_0
\end{cases}
\ ,
\end{equation}
where $m_0=0.5$.  This IMF is normalized to unity to determine the normalization constant $N_0$.  The GRB rates for the standard and narrow GRB populations (including all GRBs, not just those pointed at the Earth) are determined by Equation (\ref{dotnco}) with $K_n$ for both populations determined from the results above.  The results are shown in Figure \ref{logNlogS_rates}.  Both GRB rates are safely below the CCSNe.  The narrow jet GRB rate is a factor of 4 larger than the standard GRB rate at low redshift.

\section{Implications}
\label{implication}

In this section we explore some of the implications of a new population of nearby, narrow jetted GRBs implied by 221009A, as described in Section \ref{logNlogS_section}. 

\subsection{Finding GRBs like 221009A with gamma-ray detectors}

Assuming the model presented in Section \ref{distribution_section} is correct, and there are two populations of GRBs, in the future we would expect to detect more GRBs from the narrow jet population with GRB detectors than if this population did not exist.  We wish to determine on average how long until it is possible to rule out a model where the population is only made up of the standard GRB population, $N_1(>S_1)$, with GRB detectors.   Below we refer to this model as Model 1.  The model where there is an additional narrow jet population (hereafter referred to as Model 2) makes different predictions for the rate of GRBs above $\ga10^{-2}\ \erg\ \cm^{-2}$ than Model 1; see Figure \ref{logNlogS_fig2}.  Since almost any existing or planned GRB detector will be able to observe bursts with fluence $\ga10^{-2}\ \erg\ \cm^{-2}$, we need not concern ourselves with GRBs missed due to being below the detector's threshold.

According to \citet{burns23}, there has been $t_0=44.3$\ years of $4\pi$\ srad-equivalent observations of GRBs with existing detectors.  In that time there has been one GRB with fluence above $10^{-2}\ \erg\ \cm^{-2}$, GRB 221009A.  If Model 1 is correct, the $4\pi$\ srad equivalent total number of GRBs one would expect to be observed above $S_1$---including 221009A---is $N_{\rm obs,1}(t) = N_1(>S_1)\times (t-t_0) + 1$.  The $4\pi$\ srad equivalent number of GRBs expected under Model 2 is $N_{\rm obs,2}(t) = (N_1(>S_1)+N_2(>S_1))\times (t-t_0) + 1$, where again the $1$ includes GRB 221009A.  Model 1 can be ruled out at $n\sigma$ when
\begin{flalign}
\label{criterion}
N_{\rm obs,1}(t) < \lambda_{\rm 2,LL,n}(t)
\end{flalign}
where $\lambda_{\rm tot,LL,n}(t)$ is the Poisson lower limit on $N_{\rm obs,2}(t)$ at $n\sigma$, computed following \citet{gehrels86}.  Solving Inequality (\ref{criterion}) for $t$ as a function of $S_1$, we find that Model 1 can be ruled out at the $3\sigma$ level after a minimum of 120 years, when Model 2 predicts $N_{\rm obs,2}(t)=1.5$\ GRBs with fluence $>0.14\ \erg\ \cm^{-2}$, and Model 1 predicts only $N_{\rm obs,1}(t)=5\times10^{-2}$\ GRBs above this fluence.  Essentially this means observing one additional GRB with fluence greater than $0.14\ \erg\ \cm^{-2}$ will rule out Model 1 at $3\sigma$, and that will happen with expectation value 120 years from now if Model 2 is correct.  
Naturally the number of GRBs that will actually be observed is random.  It could take longer or shorter than 120 years to observe a second bright GRB and rule out Model 1.  Similarly it will take an expected 720 years to rule out Model 1 at the $5\sigma$ level, when Model 2 predicts $N_{\rm obs,2}(t)=13.7$ GRBs above $S_1=0.04\ \erg\ \cm^{-2}$, compared with $N_{\rm obs,1}(t)=2.15$\ GRBs for Model 1.  

Confirming the existence of the narrow jet population of GRBs from observations with $\g$-ray detectors will likely require more time than the careers of most astrophysicists.  We therefore turn to other observational consequences, in the hopes of finding another way to empirically confirm or rule out this model.  If off-axis GRBs from the narrow jet population are missed by $\g$-ray detectors, their afterglows may still be observable.



\subsection{Detection of optical afterglows by Rubin}
\label{rubinsection}

This model predicts the narrow jet GRBs will be more plentiful in the nearby Universe (Figure \ref{logNlogS_rates}), and narrower jets imply a greater number of unobserved off-axis GRBs for a given observed GRB.  As the blast wave creating an afterglow slows with time, the jet widens allowing the afterglow of an off-axis GRB to become more visible.  Thus, one might think that the narrow jet GRBs will produce more orphan afterglows than the standard GRBs.  This could potentially predict a large number of GRB afterglows detectable by the Vera Rubin Observatory.  In the $r$ band Rubin will see down to magnitude $24$ (corresponding to a flux density of about 1 $\mu$Jy) in each 30 s exposure, with a field of view of 9.6 deg$^2$ \citep{bianco22}.  It will observe approximately 3300 deg$^2$ per night \citep{ghirlanda15}.  For our purposes here, we define an orphan afterglow as a GRB afterglow that is observed without needing to be triggered by a $\g$-ray detector.  It thus includes GRB afterglows that may have been detected by $\g$-ray detectors also, but seen as afterglows without having been triggered by the $\g$-ray detectors.

\begin{figure}
\vspace{2.2mm} 
\epsscale{1.1} 
\plottwo{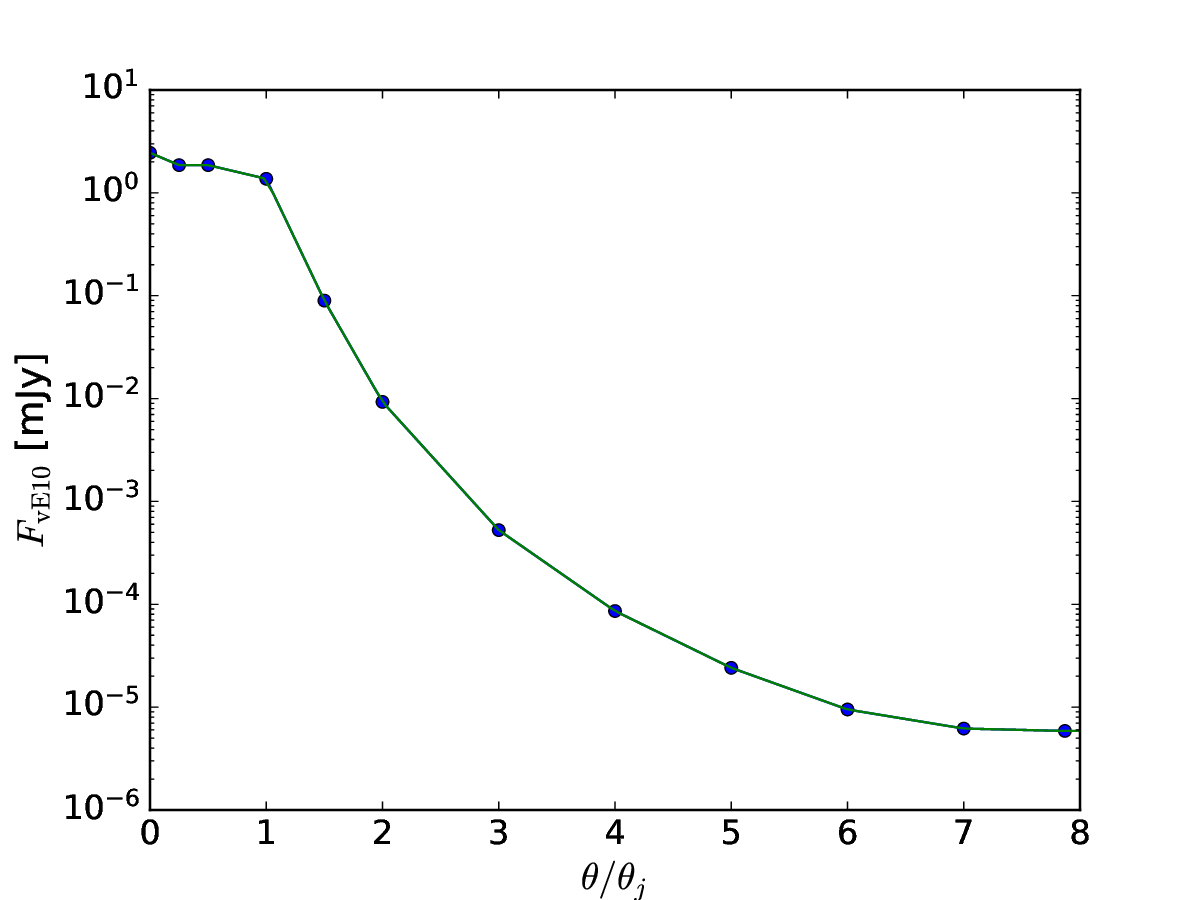}{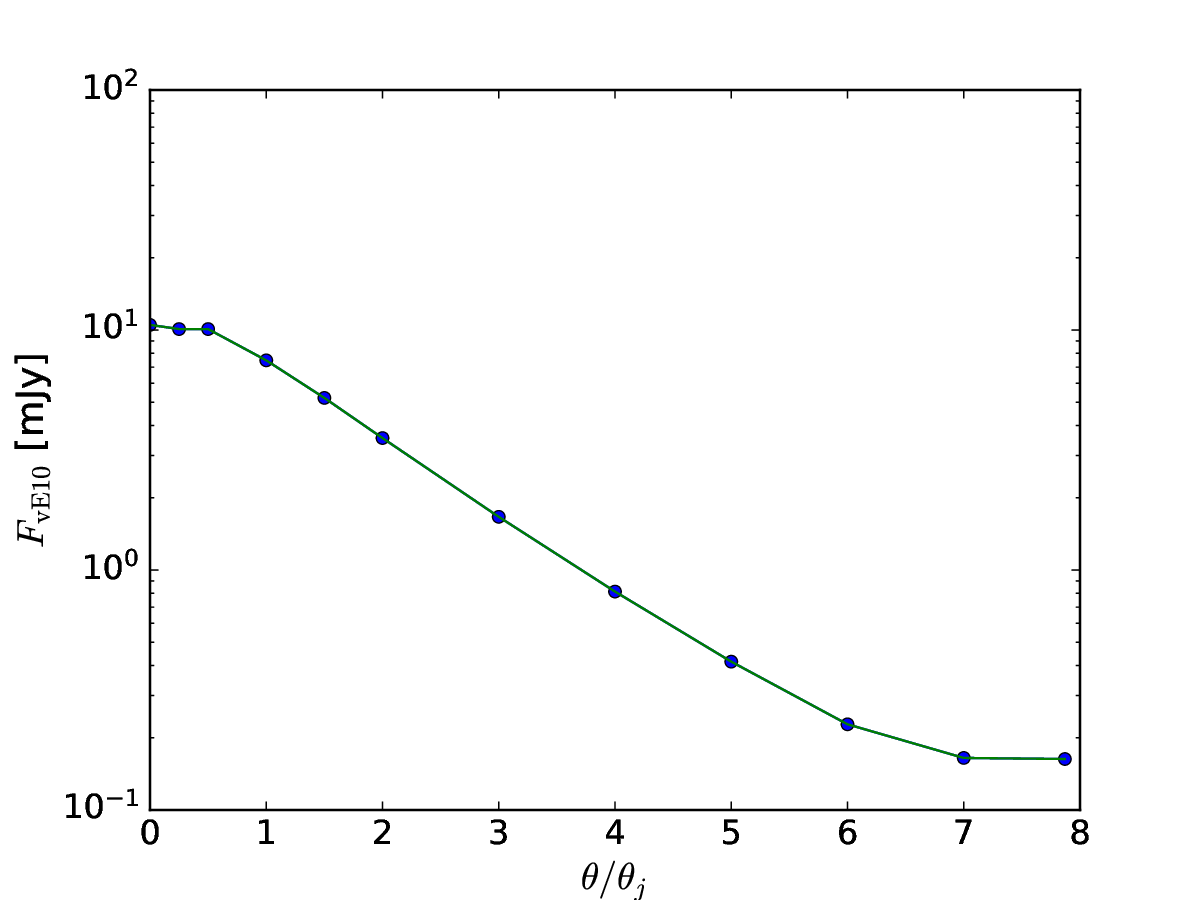}
\caption{The peak afterglow flux density as a function of $\theta/\theta_j$ for $E_\g=10^{51}$\ erg, $d_L=10^{28}$\ cm, taken from \citet{vaneerten10}.  The blue circles indicate the results taken from their light curve calculation.  The green lines indicate our interpolation.  {\em Left}:  optical ($10^{14}\ \Hz$); {\em Right}:  radio ($10^9\ \Hz$).}
\label{afterglow_flux_fig}
\vspace{2.2mm}
\end{figure}

To estimate the number of orphan afterglows, we used the afterglow modeling from \citet{vaneerten10} for their model parameters $E=10^{51}\ \erg$, $\theta_j=0.2$\ rad, $d_L=10^{28}\ \cm$, $\epsilon_B=\epsilon_e=0.1$, $p=2.5$, and a constant density circumburst medium with density $n_e=1\ \cm^{-3}$.  We consider the afterglow to be detected if at any time its flux is greater than 1 $\mu$Jy.  Therefore we are interested in the maximum flux the light curve reaches.  
For the $10^{14}$\ Hz afterglows, we find the brightest value the afterglow reaches for various values of $\theta$.  We take this flux density to be a function of $\theta/\theta_j$, which we call $F_{\rm vE10}(\theta/\theta_j)$.  For the points in between values of $\theta$, we do a simple linear interpolation.  For $\theta/\theta_j\ga8$, we set $F_{\rm vE10}$ equal to the value at $8\theta/\theta_j$.  This function is plotted in Figure \ref{afterglow_flux_fig}.  Then we assume the maximum flux density of the afterglow scales with energy and distance as the following
\begin{flalign}
\label{hatflux}
\hat F = \left(\frac{E_\gamma}{10^{51}\ \erg}\right) \left(\frac{d_L}{10^{28}\ \cm}\right)^{-2}
F_{\rm vE10}(\theta/\theta_j) 10^{-0.4A_r},
\end{flalign}
where $A_r$ is the amount of extinction (in magnitudes) by dust.

We can then calculate the log(N)-log(F) distribution for optical afterglows in a very similar way to the log(N)-log(S) distribution above.  In this case, the GRB afterglow rate is given by
\begin{flalign}
\label{dotNflux}
\frac{d\dot N}{d\Omega dF dz} = \frac{\dot n_{\rm co}(z)}{1+z} \frac{dV_{\rm co}}{dz}  \delta(F-\hat F)  \ .
\end{flalign}
The number of optical afterglows with flux density greater than $F_1$ is given by
\begin{flalign}
\label{Nflux}
N(>F_1) = \int_{F_1}^\infty dF  \int dt \int_0^1 d\mu_j g(\mu_j) \int_{\mu_j}^{1} d\mu \int dE_\gamma h(E_\gamma) 
\int^{A_{\max}}_{A_{\min}} dA_r p(A_r) 
\int_{0}^{z_{\rm max,*}} dz \int d\Omega
\frac{d\dot N}{d\Omega dS dz}\ ,
\end{flalign}
where $p(A_r)$ is the distribution of $A_r$, normalized to unity.  We use $A_{\min}=0$ and $A_{\max}=10$.  

In a study of dust extinction in GRBs, \citet{covino13} found 50\% have $V$-band extinction between $A_V\la 0.3 \div 0.4$; 77\% had $A_V<2$; and 13\% had $A_V>2$.  We choose a distribution for $A_r$ consistent with this:
\begin{equation}
p(A_r) = \begin{cases}
m_1A_r + b & 0 \le A_r < A_0 \\
m_1A_0 + b + m_2(A_r-A_0) & A_0 \le A_r < A_1 \\
m_1A_0 + b + m_2(A_1-A_0) + m_3(A_r-A_1) & A_1 \le A_r < A_2
\end{cases}\ ,
\end{equation}
where $A_0=0.35$, $A_1=2.0$, $A_2=10.0$, $b=2.4412$, $m_1=-5.7862$, $m_2=-0.232415$, and $m_3=-4.0625\times10^{-3}$.
For calculations that neglect dust extinction, we use the Dirac $\delta$ function, i.e., $p(A_r)=\delta(A_r)$.  Differences between $V$ and $r$ band extinction should not be significant, considering the precision of the estimates presented here.

Substituting Equation (\ref{dotNflux}) into Equation (\ref{Nflux}) and performing some of the integrals gives
\begin{flalign}
\label{Nflux2}
N(>F_1) & = \Delta t K_n \int_0^1 d\mu_j g(\mu_j) \int_{\mu_j}^{1} d\mu \int dE_\gamma h(E_\gamma) 
\int^{A_{\max}}_{A_{\min}} dA_r p(A_r) 
\nonumber \\ & \times
\int_{0}^{\min(z_{\max},z_{\rm max,*})} dz\ 
\Omega(F_1,\theta/\theta_j,z,E_\g,A_r)\ 
\frac{\psi(z)}{1+z} \frac{dV_{\rm co}}{dz}\ ,
\end{flalign}
where now $z_{\max}$ is found by solving
\begin{flalign}
\left(\frac{d_L(z_{\max})}{10^{28}\ \cm}\right)^2 = \frac{1}{F_1} \left(\frac{E_\gamma}{10^{51}\ \erg}\right) F_{\rm vE10}(\theta/\theta_j)10^{-0.4A_r}\ 
\end{flalign}
numerically.  The solid angle observed is
\begin{flalign}
\label{omegaeqn}
\Omega(F_1,\theta/\theta_j,z,E_\g,A_r) = \omega_{\rm Rubin} \times \frac{T(F_1,\theta/\theta_j,z,E_\g,A_r)}{1+z}
\end{flalign}
where the amount of sky observed each night by Rubin is $\omega_{\rm Rubin}\approx3300\ \deg^2$/night \citep{ghirlanda15}, and $T(F_1,\theta/\theta_j,z,E_\g,A_r)$ is the number of nights the afterglow is above the threshold $F_1$.  The parameter $T(F_1,\theta/\theta_j,z,E_\g,A_r)$ is determined from the afterglow modeling by \citet{vaneerten10}, scaled as the flux in Equation (\ref{hatflux}).  Since the afterglows by \citet{vaneerten10} are computed for a duration of $10^4$ days, $T(F_1,\theta/\theta_j,z,E_\g,A_r)$ is also scaled by a factor $(365.25\ \dday/10^4\ \dday)$, appropriate for the rate of afterglows per year.  The factor $1+z$ in Equation (\ref{omegaeqn}) takes into account cosmological time dilation.

\begin{figure}
\vspace{2.2mm} 
\epsscale{1.1} 
\plottwo{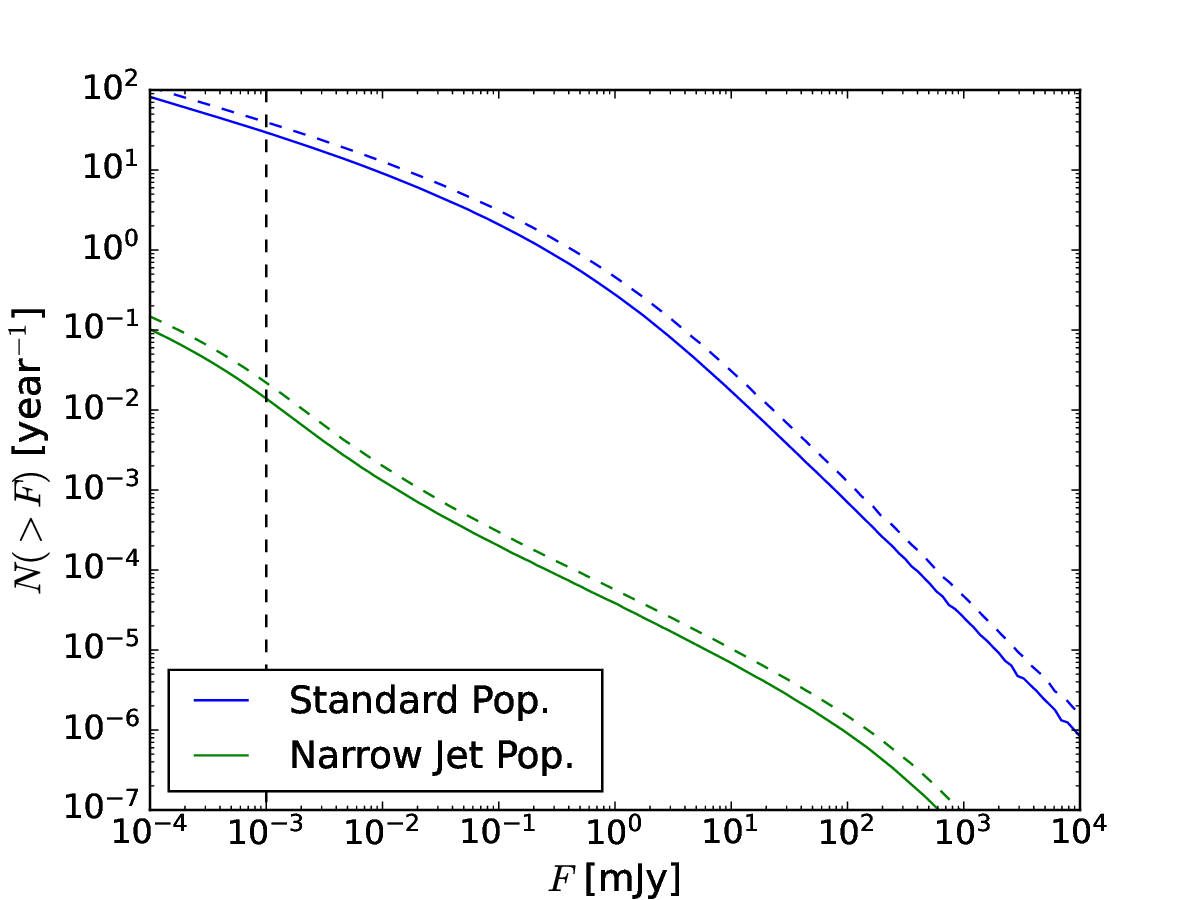}{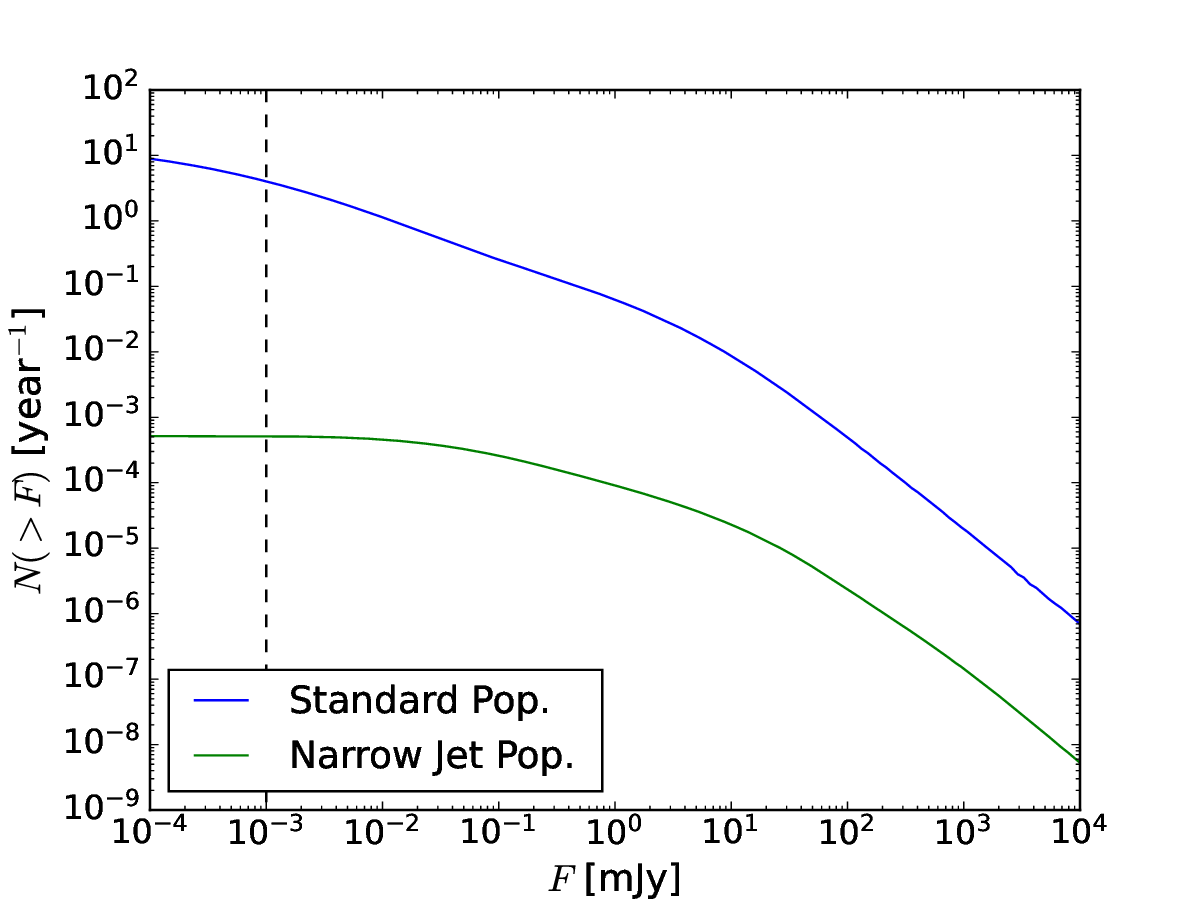}
\caption{A plot of $N(>F)$ versus flux density $F$ for the afterglows of the GRB populations with log-normal distributions for $\theta_j$ and $E_\gamma$.  {\em Left}:  the calculation for optical ($10^{14}\ \Hz$) afterglows, as would be seen by Rubin.  The solid curves include dust extinction, while the dashed curves do not.  The vertical dashed line indicates the flux sensitivity limit of Rubin for its standard 30 s exposure.  {\em Right}:  the calculation for radio ($10^9\ \Hz$) afterglows, as would be seen by SKA.  The vertical line is approximately the sensitivity limit for SKA Phase 1 used by \citet{ghirlanda14}.  }
\label{logNlogS_afterglow_fig}
\vspace{2.2mm}
\end{figure}

Using Equation (\ref{Nflux2}) we calculate the optical afterglow log(N)-log(F) distribution for both our GRB populations, using the log-normal distributions for $\mu_j$ and $E_\g$ (Equations (\ref{gmu}) and (\ref{hE}), respectively).  The result is in Figure \ref{logNlogS_afterglow_fig}.  As shown there, the number of optical afterglows from the narrow jet GRB population is negligible compared to the standard GRB population.  This simple model predicts Rubin will see about 30 standard GRB afterglows $\yr^{-1}$, the same order of magnitude as estimated by \citet{ghirlanda15}, who found Rubin would see 50 orphan afterglows $\yr^{-1}$.  It will see narrow jet GRBs at an approximate rate of only one per 100 years.  Dust extinction does not have a strong effect on our results.

We also performed the calculation setting $F_{\rm vE10}=0$ for $\theta/\theta_j>8$.  We expect this to have little effect on our results, since afterglows viewed at these large angles will indeed be quite faint, and still below the Rubin detection threshold.  And indeed the change for the afterglow log(N)-log(F) is only a small decrease in the number of sources in the narrow jet population at $F\la3\times10^{-2}$\ mJy.  Our conclusions are unchanged.



\subsection{Detection of radio afterglows by SKA}
\label{skasection}

The first phase of the Square Kilometer Array (SKA) will be complete in approximately 2028, and it should be able to see a substantial number of GRB orphan afterglows \citep[e.g.,][]{carilli04} although identification of radio sources as GRB afterglows will be challenging \citep[e.g.,][]{ghirlanda14}.  We follow the same procedure above in Section \ref{rubinsection} to estimate the number of GRB afterglows SKA is expected to see, except using the $10^9$\ Hz afterglow light curves from \citet{vaneerten10} and not including any dust extinction.  In the radio, the flux falls off at increasing $\theta$ to less than it does in the optical, as seen in Figure \ref{afterglow_flux_fig}.  We assume a sensitivity limit of 1 $\mu$Jy and the ability to observe $\omega_{\rm SKA}=1\ \deg^2$ per night, although these numbers are not totally clear.  The results for the standard population are approximately the same order of magnitude as found by \citet{ghirlanda14}.

The sensitivity and field of view of SKA are somewhat uncertain, so the normalization shown in this plot should not be taken as a firm prediction.  However, the relative number of afterglows SKA will see from the standard and narrow jet population are likely accurate (at least insofar as the model described herein is accurate).  SKA should also see $\sim10^4$ more afterglows from the standard GRB population than the narrow jet population, perhaps through commensal observations during sky surveys.


\subsection{Narrow long GRBs as the source of ultra-high energy cosmic rays}

GRBs have long been considered as candidates for the source of ultra-high energy cosmic rays \citep[UHECRs; e.g.,][and references therein]{meszaros06}.  It is natural to consider if the new population of narrow-jet GRBs proposed here could be the source of UHECRs.

Assuming that GRB host galaxies are within clusters of galaxies, the deflection angle of the highest-energy cosmic rays with energy $E_{\rm uhecr} = 100$~EeV due to random magnetic field is
\citep{Dermer2009NJPh...11f5016D}
\begin{equation}
    \theta_{\rm dfl} \simeq 5^\circ\ Z \left( \frac{B_{\rm ICM}}{0.1~\mu\rm G} \right) \left( \frac{d}{5~\rm Mpc} \right) \left( \frac{E_{\rm uhecr}}{100~\rm EeV} \right)^{-1} \left( \frac{N_{\rm inv}}{500} \right)^{-1/2} 
\end{equation}
for a nucleus of charge $Z$. Here $d = 5$~Mpc is the radius of the cluster, $B_{\rm ICM} = 0.1~\mu$G is the cluster magnetic field \citep{Carilli2002ARA&A..40..319C} with a coherence length of $\lambda\approx 10$~kpc and $N_{\rm inv} = d/\lambda$. Therefore, Fe nuclei, which are the dominant composition of UHECRs at the highest energies \citep{Auger2017JCAP...04..038A}, if accelerated in GRB jets, would be completely isotropized after propagation through galaxy clusters. As such, UHECR Fe from all GRBs taking place within the GZK radius of $d_{\rm GZK}\approx 100$ Mpc can be observed from Earth, regardless of the orientation of the jet.  If the amount of energy released in UHECR Fe nuclei from a GRB is comparable to its electromagnetic energy released, $E_\gamma \approx 10^{51}$~erg, then the rate of energy injection in UHECRs within the GZK volume from the narrow-jet population of GRBs with a local rate of $\dot n(z\approx 0) \approx 4\times 10^{-8}$~Mpc$^{-3}$~yr$^{-1}$, is $j _{\rm CR}(z\approx0) \approx 4\times 10^{43}$~erg~Mpc$^{-3}$~yr$^{-1}$. Note that an energy injection rate of $\approx (0.5\div 4)\times 10^{44}$~erg~Mpc$^{-3}$~yr$^{-1}$, for cosmic rays with energy $\gtrsim (1\div 3)\times 10^{19}$ eV, in the local Universe is needed to energize UHECRs against GZK losses \citep{Waxman1997PhRvL..78.2292W, Katz2009JCAP...03..020K}.  Therefore, the narrow jet population, with a modest ($\approx 1\div 10$) baryon loading factor, could be the sources of the highest-energy cosmic rays if they are Fe nuclei.  The energy injection rate in UHECRs from the standard population of GRBs would be much lower, $j_{\rm CR}\approx 10^{43}$~erg~Mpc$^{-3}$~yr$^{-1}$ \citep[see also][]{Dermer2010ApJ...724.1366D}. 

The evolution of the source of UHECRs with redshift is not clear.  If they are made up of intermediate mass nuclei with few protons or iron nuclei, the redshift evolution of the source of UHECRs needs to be negative if the UHECRs are accelerated with index $p\approx2$, as expected from test particle shock acceleration \citep{Taylor2015PhRvD..92f3011T}.  That is, if the number of UHECR sources per comoving volume is $n(z)\propto (1+z)^\alpha$, then $\alpha<0$.  \citet{Das2021EPJC...81...59D} found that the UHECR source can be non-evolving ($\alpha=0$) or exhibit positive evolution ($\alpha>0$) and can reproduce observations with $p\approx 2$ if there are two populations that are the sources of UHECRs, each with different chemical compositions.  Positive evolution is consistent with the narrow jet population of GRBs described here being the source of UHECRs.

\subsection{Implications for extinction events on Earth}

Based on extensive simulations by \citet{thomas05a,thomas05b}, \citet{piran14} defined the fluence of a GRB that will cause a mass extinction of life on Earth as $S_{\rm ext}\approx100\ {\rm kJ}\ \m^{-2} = 10^8\ \erg\ \cm^{-2}$.  We follow those authors and use this as the threshold to estimate how often extinction events from GRBs will occur.  GRB 221009A had a fluence of $S_{\rm BOAT}= 0.21\ \erg\ \cm^{-2}$ at a redshift $z_{\rm BOAT}=0.151$, giving it a luminosity distance of $d_L=713$\ Mpc for our assumed cosmology.  So this burst would cause an extinction event on Earth if it occurred at a distance $d_L<32\ \kpc$.  This is approximately the diameter of the Milky Way, but smaller than the distance to nearby galaxies such as M31 ($\approx$900 kpc) and the Large and Small Magellanic Clouds ($\approx 50\ \kpc$).  {\em If a GRB similar to 221009A has exploded anywhere in our Galaxy and was pointed at Earth, it would have caused a mass extinction event.}  This is farther than is generally assumed for distances where GRBs from the standard population would cause extinction events, typically a few kpc.  It is thus interesting to estimate how often events from the narrow jet population would occur in the Milky Way.  

The total Milky Way star formation rate has been measured to be $\Psi_{\rm MW} = 2.0\pm 0.7\ M_\odot\ \yr^{-1}$ by \citet{elia22}  and $\Psi_{\rm MW}=1.65\pm0.19\ M_\odot\ \yr^{-1}$ by \citet{licquia15}.  We will use $\Psi_{\rm MW}\approx 2\ M_\odot\ \yr^{-1}$ and make the simplifying assumption that the star formation rate is the same everywhere in the Milky Way and independent of distance, so that the GRB distribution with distance $x$ is $k(x)=1/x_{\max}$, where we take $x_{\max}=32$\ kpc. The
number of GRBs in the Milky Way above the extinction fluence threshold $S_{\rm ext}$ is
\begin{flalign}
\label{NMW1}
N_{\rm MW}(>S_{\rm ext}) = \int_{S_{\rm ext}}^\infty dS \int d\Omega \int dt \int_0^1 d\mu_j g(\mu_j) \int_{\mu_j}^{1} d\mu \int dE_\gamma h(E_\gamma) \int_0^{x_{\max}} dx \frac{d\dot N}{d\Omega dS}
\end{flalign}
where
\begin{flalign}
\label{dotN_MW}
\frac{d\dot N}{d\Omega dS} = K\p_n \Psi_{\rm MW} \delta(S-\hat S)\ ,
\end{flalign}
and now
\begin{flalign}
\label{fluence_MW}
\hat S = \frac{E_{\g}}{4\pi x^2 (1-\mu_j)}\ .
\end{flalign}

First, we use the model that assumes all GRBs from our narrow jet population (from which 221009A is drawn) have the same $\theta_j$ and $E_\gamma$ as we did in Section \ref{modeldeltafcn}.  In this case, Equations (\ref{NMW1}), (\ref{dotN_MW}), and (\ref{fluence_MW}) simplify to
\begin{flalign}\
\label{NMW2}
N_{\rm MW}(>S_{\rm ext}) = \Omega \Delta t K_n\p \Psi_{\rm MW}(1-\mu_j)\ .  
\end{flalign}
We can determine the normalization constant $K_n\p$ from from the overall rate of our narrow jet population, using the example of GRB 221009A.   Solving Equation (\ref{NSapprox}) with the measured fluence for GRB 221009A, $S_{\rm BOAT}=0.21\ \erg\ \cm^{-2}$, for $K\p_n$, and substituting it in Equation (\ref{NMW2}), we get
\begin{flalign}
\label{NMW3}
N_{\rm MW}(>S_{\rm ext}) = \frac{3\Psi_{\rm MW} N(>S_{\rm BOAT})}{a_s} 
\left( \frac{H_0}{z_{\rm BOAT}c}\right)^3\ .
\end{flalign}
Based on the Poisson rate for GRB 221009A, we expect $N(>S_{\rm BOAT})\ga 4\times10^{-3}\ \yr^{-1}$ (see Section \ref{modeldeltafcn}).  These values in Equation (\ref{NMW3}) give $N_{\rm MW}(>S_{\rm ext}) \ga 10^{-8}\ \yr^{-1}$.  In other words, life on Earth should experience one extinction event from a GRB from this population per 100 Myr.  This result is independent of the details of GRB 221009A, i.e., its $E_{\rm \g}$ and $\theta_j$, as shown in Equation (\ref{NMW3}).  It assumes only that GRB 221009A has the fluence and redshift that have been measured, that it occurs at a rate consistent with the Poisson rate for GRBs being detected within the time that humanity has been capable of detecting GRBs, and that the star formation rate in the nearby Universe and the Milky Way have the values that have been measured.

We can perform a more detailed estimate of the rate of extinction level GRBs using the distributions of $g(\mu_j)$ and $h(E_\gamma)$ from Section \ref{distribution_section}.  In this case, Equations (\ref{NMW1}), (\ref{dotN_MW}), and (\ref{fluence_MW}) give
\begin{flalign}
N_{\rm MW}(>S_{\rm ext}) = \Omega \Delta t K\p_n \Psi_{\rm MW} \int_{0}^{1} d\mu_j g(\mu_j)(1-\mu_j)
\int dE_\gamma h(E_\gamma) \min\left[ 1, \left\{\frac{E_\g}{4\pi S_{\rm ext}x_{\max}^2(1-\mu_j)}\right\}^{1/2} \right]\ .
\end{flalign}
For the distributions and parameters in Section \ref{distribution_section}, $N_{\rm MW}(>S_{\rm ext}) \ga 1.9\times10^{-9}\ \yr^{-1}$, or {\em one extinction level GRB per 520 Myr.}  It has been suggested that a GRB could have caused the end-Ordovician mass extinction 440 Myr ago \citep{melott04}.  It is interesting to note that the time since this extinction event is $\approx 1/N_{\rm MW}(>S_{\rm ext})$ above.  Of course, the occurrence of these GRBs in the Milky Way is random, not periodic, so one would not expect such a GRB exactly every 520 Myr.

If this model is correct, not only would the Earth experience a mass extinction level GRB with a rate of one per $\approx 500$\ Myr, but everywhere in the Galaxy would.  GRBs from the narrow jet population could cause extinction events on other planets in the Galaxy that harbor life.  Previous work has assumed GRBs are dangerous to life out to no more than $\sim 1$\ kpc.  Authors have considered a ``Galactic Habitability Zone'' where life could develop and evolve \citep[e.g.,][]{lineweaver04,gowanlock16,spinelli23}.  Life in the inner portion of the Milky Way, where star formation is greater, would be devoid of life due to supernovae and GRBs; and the outer region of the Milky Way has too low of metallicity for terrestrial planets to form.  \citet{lineweaver04} suggest there is a Galactic Habitability Zone in between these two extremes where life would be most likely to evolve.  Long GRBs are associated with lower metallicity environments; this led \citet{spinelli23} to suggest that large distances from the Galactic center where metallicity is lower may also be a difficult place for life due to more GRBs, assuming GRB rate correlates with metallicity.  However, if everywhere in the Milky Way experiences an extinction event on average every 500 Myr from GRBs like 221009A, the Galactic Habitability Zone must be re-evaluated.  The Milky Way could be a more sterile place than previously thought.  This could have implications for the search for life in the Milky Way in the future with NASA's Habitable Worlds Observatory (HWO).







\section{Discussion}
\label{discussion}

We have developed the hypothesis that GRB 221009A, the BOAT, is from a population of nearby, narrow jet GRBs separate from most other long GRBs that are observed.  This separate distribution of GRBs with the fluence greater than that of 221009A occur roughly once per 200 years, rather than every 10,000 years \citep[as estimated by][]{burns23}, and all GRBs from this population occur approximately one per 30 years.  This makes this burst more consistent with the fact that it was observed within the time period that humanity has had the ability to observe GRBs, and makes the observed GRB rate for this population comparable to that of Galactic supernovae.

Verifying this hypothesis will be challenging and time consuming, considering how rarely they occur pointing in our direction.  To detect the narrow jet population of bursts, one needs detectors that can see as much of $4\pi$ srad of the sky as possible, operating for a long time--likely more than 100 years.  Sensitivity is less important, since these GRBs will be so bright that almost any $\g$-ray detector will be able to observe them.  Indeed, their brightness may face its own observational challenges, since 221009A saturated most $\g$-ray detectors.  More important is coverage in solid angle and time.  The proposed MoonBEAM \citep{fletcher23} would be in a lunar resonance orbit and would observe nearly the entire sky at any given time.  This is the sort of mission that would be ideal, and ones like it will likely need to be flown for the next few hundred years in order to confirm this new population of GRBs.  Observatories such as Rubin and SKA are unlikely to see enough orphan afterglows to confirm this population, as we show in Sections \ref{rubinsection} and \ref{skasection}.  

It is unclear why there would be a separate population of long GRBs at low redshift.  Several authors \citep{yu15,petrosian15} studied the redshift evolution of the luminosity function of long GRBs detected by {\em Swift} and found an excess of GRBs at low redshift relative to the cosmic star formation rate.  \citet{le17} found a similar low-redshift excess with a slightly different method.  However, \citet{yu15} mention that a selection effect based on the fact that not all Long GRBs have measured redshifts could be the origin of their low-$z$ excess.  Further work by \citet{le20} with an updated GRB sample found no low redshift excess, indicating there was a selection effect in the previous sample used by these authors.  Curiously, \citet{ghirlanda22} found the opposite---{\em fewer} GRBs at low-$z$ relative to the cosmic star formation rate--using BATSE and {\em Fermi}-GBM detected bursts.  All these results were prior to the discovery of GRB 221009A.  Assuming the rate for GRBs like 221009A are consistent with its observation in the time GRB detectors have existed, {\em we have shown that a low-$z$ excess in the rate of GRBs is inevitable.}

One possible explanation proposed for the excesses found by \citet{yu15} and \citet{petrosian15} is the evolution in the jet opening angle of GRBs.  \citet{lloyd19,lloyd20} estimate that jet opening angle {\em decreases} with increasing redshift. This could affect star formation rate estimates from Long GRBs \citep{lloyd20_sfr}.  However, here we find the opposite:  we show that the low-redshift GRB population (from which 221009A originates) {\em must have smaller opening angles than the standard GRB population}.

Another possible explanation is that there is a population of low-$z$ Long GRBs that are from compact object mergers, as suggested by \citet{petrosian23}.  This was inspired by the detection of 230307A, a Long GRB that shows evidence for a kilonova \citep{dichiara23}.  However, for 221009A, the light curve of a core-collapse SN seems to be emerging \citep{fulton23}, so this is an unlikely origin for this GRB and its parent low redshift population.  

We speculate that there could be a population of Long GRBs that originate in higher metallicity environments only found in the nearby Universe, and that this prefers narrower jets than the standard population of GRBs.  If the progenitors of the nearby population are metal-rich stars, this could lead to greater mass loss for the stars, and hence less material for a narrow jet to penetrate through the remaining stellar envelope.  Modeling is necessary to determine if this is at all plausible.  However, we note that the host galaxy of GRB 221009A does seem to have metallicity consistent with other Long GRBs \citep{malesani23}, perhaps making this explanation less likely.


The narrow-jet population of GRBs occurring in galaxy clusters can potentially supply the energy needed for the observed diffuse flux of UHECRs above $\sim 10^{19}$ eV because of their higher intrinsic rate than the standard GRB population in the nearby universe. This, however, requires that heavy nuclei such as Fe are accelerated in GRB shocks, which might be possible in some scenarios \citep[see, e.g.,][]{Wang2008ApJ...677..432W}. We estimate that a moderate, of the order of a few, baryon loading is needed for the narrow-jet population to be the sources of UHECRs as compared to a few tens for the standard GRB population.

The existence of a nearby population of bright GRBs also has implications for the rate of GRBs that can cause mass extinctions of life on Earth.  The GRB 221009A, with the highest $E_{\rm iso}$ of any observed burst, is bright enough that it could cause an extinction event on Earth if it occurred anywhere in the Milky Way.  This is contrary to what is usually assumed for the distances where a GRB could cause an extinction event typically $\sim\kpc$ \citep[e.g.,][]{piran14}.  We show that, if GRB 221009A occurred at the rate consistent with being detected in the timescale of when GRBs have been detectable by humanity, a GRB extinction event on Earth from this population should occur at a rate of approximately one per 500 Myr.  The result is mostly independent of the details of the GRB population.  It is a fairly simple estimate, and is likely accurate within a factor of a few.  A more detailed estimate would take into account the geometry of Milky Way and differences in star formation in different parts of the Galaxy.  This rate is consistent with the timescale of the end-Ordovician mass extinction event, which has been hypothesized to be caused by a GRB.  These extinctions also have implications for astrobiology and the search for life on exoplanets by future observatories such as HWO. Determining the rate of GRBs like 221009A is essential for understanding how common or rare life is in the Milky Way.

\begin{acknowledgements}

J.D.F.\ would like to acknowledge conversations with Justin Vandenbroucke that led to the work presented in this paper, and Matthew Kerr for suggesting a calculation of the afterglows detectable by SKA.  The authors would like to thank Eric Burns for the logN-logS data plotted here, and for useful discussions.  J.D.F.\ is supported by NASA through contract S-15633Y, and by the Office of Naval Research.  S.R.\ is supported by grants from the National Research Foundation (NRF), South Africa,  BRICS STI programme; the National Institute for Theoretical Computational Sciences (NITheCS), South Africa; and the South African Gamma-ray Astronomy Programme (SA-GAMMA).

\end{acknowledgements}

\bibliography{grb_ref}{}
\bibliographystyle{apj}


\end{document}